\documentstyle[12pt]{article}
\newskip\humongous \humongous=0pt plus 1000pt minus 1000pt
\def\caja{\mathsurround=0pt}

\def\eqalign#1{\,\vcenter{\openup1\jot \caja
        \ialign{\strut \hfil$\displaystyle{##}$&$
        \displaystyle{{}##}$\hfil\crcr#1\crcr}}\,}
\newif\ifdtup

\def\be{\begin{equation}}
\def\ee{\end{equation}}
\def\ba{\begin{eqnarray}}
\def\ea{\end{eqnarray}}

\begin{document}
\renewcommand{\theequation}{\thesection.\arabic{equation}}
\newcommand{\beq}{\begin{equation}}
\newcommand{\eeq}[1]{\label{#1}\end{equation}}
\newcommand{\ber}{\begin{eqnarray}}
\newcommand{\eer}[1]{\label{#1}\end{eqnarray}}
\begin{titlepage}
\begin{center}

\hfill CERN-TH/96-153   \\
\hfill hep-th/9611205\\

\vskip .5in

{\large \bf $F^4$ TERMS IN $N=4$ STRING VACUA $\ ^{\dag}$}
\vskip 0.1in

\vskip .5in

{\bf C. Bachas$ \ ^{\spadesuit}$
and E. Kiritsis$ \ ^{\clubsuit}$}
\vskip .1in

{
$\ ^{\spadesuit}$
 Centre de Physique Th\'eorique, Ecole Polytechnique,
 91128 Palaiseau, FRANCE \\
 email: {\it bachas@orphee.polytechnique.fr}\\
\   \\
$\ ^{\clubsuit}$
 Theory Division,
CERN, CH-1211, Geneva 23, SWITZERLAND\\
email: {\it kiritsis@nxth04.cern.ch}
}

\vskip .15in

\end{center}

\vskip .4in

\begin{center} {\bf ABSTRACT }
\end{center}

\vskip .15in

\begin{quotation}

\noindent

We discuss  $F_{\mu\nu}^4$ terms in torroidal
 compactifications of type-I and heterotic $SO(32)$ string theory.
 We give a simple argument
 why only short BPS  multiplets contribute to these terms at one
loop, and verify heterotic-type-I duality to this order.
Assuming exact duality,  we exhibit in the heterotic calculation
non-zero terms that are two-loop, three-loop and non-perturbative
on the type-I side.

\end{quotation}
\vskip 1in
CERN-TH/96-153\hfill\\
November 1996\hfill\\
\vfill
{$\ ^{\dag}$ To appear in the Proceedings of the Trieste Spring
School and
Workshop in String Theory, April 1996}

\end{titlepage}
\vfill
\eject
\def\baselinestretch{1.2}
\baselineskip 16 pt
\noindent
\setcounter{equation}{0}

\setcounter{equation}{0}
{\it Introduction}.
BPS states play a special role  in
theories with extended ($N\geq 2$) supersymmetry.
The fact that they form multiplets which are shorter than the generic
representation of the supersymmetry algebra implies relations
between their mass, charges and values of moduli which are  valid
in the exact  quantum theory. For  $N\geq 4$ these relations are
furthermore  purely classical, and they  ensure that BPS states are
either stable or, at worse,  marginally-unstable. Stable BPS states
can thus be traced all the way to strong coupling, and their
existence with appropriate multiplicities
 has constituted  the main test of
the various  duality conjectures.

  Another remarkable feature of  BPS states is that they
saturate certain one-loop terms in the effective low-energy
action.  This fact has been  articulated
 clearly by Harvey and Moore \cite{HM}
in the context of heterotic $N=2$ thresholds, though it was
implicit in much of the earlier work, such as for instance
refs.  \cite{L}.
 BPS-saturated terms are furthermore typically related,
 by supersymmetry, to anomalies, and  are thus expected to obey
non-renormalization theorems. This  makes  them  a precious tool for
checking  duality conjectures. Tseytlin \cite{Ts}
has in particular used such $F_{\mu\nu}^4$ terms, in order
to  test the  conjectured duality between the type-I and
heterotic-string theories in ten dimensions \cite{W,WP}.
In this paper we will extend Tseytlin's analysis to torroidal
compactifications.

The effective gauge-field action of open-string theory
is closely related to the phase-shift or velocity-dependent
forces between D-branes \cite{D,Lif,GG,Sun,DKPS}. BPS saturation
 and a non-renormalization assumption of the
leading $o(v^4)$ interaction are, furthermore, a crucial
ingredient in the recent interesting conjecture by Banks et al
\cite{M} concerning M-theory in the infinite-momentum frame.
Despite their close relation the two calculations differ
however in some  significant ways. For instance
 in  the effective-action calculation
one  subtracts  diagrams with massless closed strings in the
intermediate channels. These diagrams  must be kept in the D-brane
calculation, where they  are regulated effectively
by the world-volume  dimensional reduction. Our analysis
does  not therefore translate  into the D-brane context immediately,
but it raises by analogy some interesting questions.

\vskip 0.3cm

{\it Supertrace formulae}.
BPS saturation at one loop follows from  supertrace
formulae \cite{Fer} involving
 powers of helicity and R-symmetry charges.
These  are  easier  to discuss in terms of generating
functionals. Define  for instance
$$
Z_{rep}(y) = str\  y^{2 \lambda} \eqno(1)
$$
where the supertrace stands for a sum over bosonic minus fermionic
states of  the representation, and $\lambda$ is
the eigenvalue of a generator of the little
group: $SO(3)$ or $SO(2)$ in the massive, respectively massless case
in four dimensions.
For a particle of spin $j$ we have
$$
Z_{[j]} = \cases{& $(-)^{2j} \Bigl( { y^{2j+1}-y^{-2j-1}
\over y- 1/y } \Bigr)$ \ \  {\rm massive} \cr
 &\ \cr
 & $(-)^{2j} ( y^{2j}+y^{-2j})$ \ \ \ \
  {\rm massless} \cr}
 \eqno(2)
$$
When tensoring representations the generating functionals get
multiplied,
$$
Z_{r\otimes {\tilde r}} = Z_r Z_{\tilde r} \ . \eqno(3)
$$
The supertrace of the $n$th power of helicity can be extracted
from the generating functional through
$$ str\ \lambda^n =   (y^2{d\over dy^2})^n \ Z(y)\vert_{y=1} \ .
\eqno(4)
$$

 Consider now $N=2$ multiplets. The supersymmetry
algebra contains four  fermionic charges that may
 act independently: two
of them raise the helicity  by one half unit,
 while the other two lower it by the same amount.
 For the generic massive (long)
multiplet
all charges act non-trivially  on some ``ground state'' of spin $j$
and one finds
$$
Z_{long}^{N=2} = Z_{[j]}\ (1-y)^2(1-1/y)^2  \ \ . \eqno(5a)
$$
For a massless or a
short massive multiplet half of the supercharges have a trivial
action so that  one  finds instead
$$
Z_{ short}^{N=2} = (2)\times  Z_{[j]}\ (1-y)(1-1/y)   \ \ , \eqno(5b)
$$
where the factor 2 is required in the massive case, since short
massive
multiplets carry charge and  are thus  necessarily complex.
 Familiar examples of short massive multiplets include the
monopoles ($j = 0$) and charged gauge bosons ($j={1\over 2}$)
of pure  N=2 Yang-Mills theories. An immediate consequence of eqs.
(5)
is that {\it only for short (BPS) multiplets is}
 $str\ \lambda^2 \not= 0$.

  Let us turn next to the $N=4$ algebra. This has   four raising and
four lowering fermionic charges, all of which can  act
independently  in  a  generic massive (long) representation,
$$
Z_{long}^{N=4} = Z_{[j]}\ (1-y)^4(1-1/y)^4  \ \ . \eqno(6a)
$$
Short representations, which include all the massless as well as
some massive  multiplets, annihilate half the supercharges so that
$$
Z_{ short}^{N=4} = (2)\times Z_{[j]}\ (1-y)^2(1-1/y)^2   \ \ .
\eqno(6b)
$$
This is precisely the content of a long $N=2$
representation.
$N=4$ has also  intermediate (or semi-long) multiplets,
which annihilate  one-quarter of supercharges, and for which
$$
Z_{semi-long}^{N=4} = 2\times Z_{[j]}\ (1-y)^3(1-1/y)^3  \ \ .
\eqno(6c)
$$
The factors of two take again into account that
  massive short and intermediate multiplets have charge
and are thus necessarily complex.
It follows trivially from the above expressions that $str \lambda^2
=0$
always,  $str \lambda^4 \not= 0$ only for short multiplets, and
$str \lambda^6 \not= 0$ in  both the short and the intermediate case.

The  discussion can be extended easily to take into account
 R-symmetry
charges. These  are simply   helicities in some  (implicit) internal
dimensions: there is a single R-charge for $N=2$, and three
independent
charges, corresponding to the Cartan generators of $SO(6)$, in the
 $N=4$ case. To get a non-zero result for short, intermediate or
long multiplets in the latter case, one must insert in the supertrace
at least four, six, respectively eight powers of helicity and/or
R-charges.

\vskip 0.3cm

{\it Type-I effective action}.
 Let us turn now to  the one-loop calculation
of the effective gauge-field action in type-I theory. In the
background-field method the one-loop free energy
in $d$ non-compact dimensions  reads  \cite{BP}
$$\eqalign{
{\cal F}_I^{(1)}(B) =& -{V^{(d)} \over 8\pi} \int_0^\infty
{dt\over t}
 (2\pi^2 t)^{1-{d\over 2}} \times\cr\times
& Str\  {q{\cal B}\over {\rm sinh}(\pi t\epsilon/2)}\   e^{-{\pi
t\over 2}
(M^2+ 2\lambda\epsilon)} \cr} \eqno(7)
$$
where $B={\cal B}Q$ is a  background magnetic field
pointing in some direction $Q$ in group space, $q=q_L+q_R$\  is
the
corresponding charge  distributed between  the two string
endpoints, and $\epsilon$ is a  non-linear
function of the charges and the field that vanishes linearly
with the latter
$$
 \epsilon ({\cal B},q_L,q_R)  \simeq q {\cal B} + o({\cal B}^3) \ .
\eqno(8)
$$
In the weak-field limit and for low spins this is a familiar
field-theory expression: it follows directly from the fact that
elementary  charged particles have gyromagnetic
ratio 2 and a spectrum given by equally-spaced Landau levels.
 The effects of non-minimal coupling for
 an open string are  captured
essentially by the replacement
 $q {\cal B}\rightarrow \epsilon$.

 The supertrace  in eq. (7)
 runs over all charged string states. For any
given supermultiplet
 the mass and charges  are however common, so that
its  contribution is proportional to
$$ str \ e^{-\pi t \epsilon\lambda}
 = \sum_{n=0}^{\infty} {(\pi t\epsilon)^{2n}\over (2n)!}\
 str \lambda^{2n} \ , \eqno(9)
$$
where we have used the fact that odd powers of helicity trace out
automatically to zero.
Since the $\epsilon$-expansion is an expansion in  weak-field, the
 various non-renormalization statements
at one loop follow directly from the properties of
 helicity supertraces and eqs. (7,9).
Thus in $N=2$ theories
 the first non-zero term,  proportional to
$ str\lambda^2$, is the one-loop gauge kinetic function:
it only  receives contributions
 from short (BPS) multiplets, as has been  noted previously
by  using identities of
$\theta$-functions \cite{DL,BF}.
In  $N=4$ theories the gauge coupling constant is
 not  corrected at one loop.
The first non-zero term, proportional to $ str\lambda^4$
is  quartic in the background field, and
only receives contributions from short $N=4$ multiplets. This was
 noted
again through $\theta$-function identities in
the D-brane context in  refs. \cite{DKPS,GG}.
The following term of order $o(F^6)$
 is also determined, incidentally,
by short BPS states. This is because long multiplets do not
contribute to $str \lambda^6$, and there are no
intermediate multiplets in the perturbative type-I spectrum.

  Let us take now a closer look at the quartic term arising
in N=4 (torroidal) compactifications. The only
perturbative charged  BPS states  are the multiplets of the
$SO(32)$ gauge bosons, together with all  their Kaluza-Klein
descendants.
For these states the  mass is equal to the internal momentum, so that
after some straightforward algebra  one finds
$$\eqalign{
{\cal F}_I^{(1)}/& V^{(d)}  = - { {\cal B}^4\over 2^9 \pi^4}
\int_0^\infty {dt\over t} (2\pi^2 t)^{4-{d\over 2}}
\times\cr \times
& \sum_{\rm
Chan\atop Patton}
(q_L+q_R)^4
\sum_{p\in a_L+a_R+^*\Gamma}  e^{-\pi t p^2/2}
 + o({\cal B}^6)  \cr} \eqno(10)
$$
where  $\ ^*\Gamma$ stands for the
$(10-d)$-dimensional  lattice of Kaluza-Klein momenta,
which must be shifted from the origin in the presence of
non-vanishing Wilson lines. Each  end-point charge  takes
32 values, but the sum runs only over antisymmetric states.
For ease of notation we will from now on
suppress the $o({\cal B}^6)$ terms when writing effective actions.

The above expression
 is strictly-speaking  formal, since  it diverges
at the $t\rightarrow 0$ limit of integration.
This is an open-string ultraviolet  divergence, but can be also
interpreted as coming from  an
on-shell dilaton or  graviton that propagates  between two
non-vanishing
tadpoles. We are  interested in the effective
 (\"Wilsonian\") action,
so this divergence due to exchange of  massless particles
must  be subtracted away.
The right procedure is to change variables to the closed-string
proper time $l$, which is related to $t$ differently
 for the annulus and M\"obius-strip
contributions,
$$ l = \cases{ $1/t$\ &{\rm annulus }\cr
$1/4t$\ & {\rm M\"obius strip}\cr} \eqno(11)
$$
Separating the two topologies amounts to writing the sum over
Chan-Patton states as an unconstrained sum over all left- and
all right- endpoints, minus the diagonal. After performing also a
Poisson resummation  the result reads
$$\eqalign{
\ \ \ \ \ {\cal F}_I^{(1)}  = - { {\cal B}^4 V^{(10)} \over 2^{10} \pi^6}
\int_0^\infty dl &
\ {1\over 2}\sum_{w\in\Gamma} \times 
 \Bigl\{ 
\sum_{\vert L>,\vert R>} (q_L+q_R)^4
  e^{-w^2 l/2\pi + iw\cdot(a_L+a_R)} -
\cr - & 4\times
\sum_{\vert L>=\vert R>} (2q_L)^4
 e^{-2w^2 l/\pi + 2iw\cdot a_L}
\Bigr\}\cr} \eqno(12)
$$
where $\Gamma$ is now the compactification lattice.
Our conventions are such that $w=2\pi R m$ for a circle of radius
$R$.

The divergence  in the above expression comes from the $w=0$ piece,
as  all other terms  are  exponentially-small  in the
$l\rightarrow\infty$ region. Thanks to the factor $4$ that multiplies
the M\"obius-strip contribution, this divergence is proportional
to  $(tr B^2)^2$. It
 corresponds {\it precisely} to the
 tadpole $\rightarrow$ massless-propagator$\rightarrow$
tadpole  diagram,  that must
be removed in the  effective action
 \cite{BF}.
Switching-off the Wilson-lines for simplicity,
and changing integration variable once again for the M\"obius
contribution,
 we thus obtain our final expression
$$
 \int {\cal L}_I^{(1)} =
- { V^{(10)} \over 2^{10} \pi^6}\  \{ 24\  tr B^4 +  3 (tr B^2)^2 \}
 \times \int_0^\infty dl
\sum_{w\in\Gamma-\{0\}}
  e^{-w^2 l/2\pi}
 \eqno(13)
$$

Since in the decompactification limit  all $w\not= 0$ terms
disappear,
we have just shown in particular
 {\it  that the 10d effective type-I Lagrangian has no
one-loop}
$F^4$ {\it corrections}. This is in agreement with 10d
heterotic-type-I
duality \cite{Ts}
as we will discuss in detail  in the following section.
The fact that only
 open  BPS states contribute  to the amplitude  is
in this respect crucial:
it ensures  that the string-scale does not enter in
the expression for  ${\cal F}_I^{(1)}$,
which must  thus cancel entirely
when passing to  the effective
action in ten dimensions.  More generally,
after compactification, the fact that  ${\cal F}_I^{(1)}$
does not depend on  $\alpha^\prime$
 implies that all corrections to the
effective Lagrangian at one loop  come  from integrating out the
Kaluza-Klein modes of massless 10d string  states.

\vskip 1.0 cm

{\it Heterotic-type-I duality}.
The predictions of this string-string duality \cite{W,WP}
for the effective action in 10d,
have been worked out
and checked against earlier  calculations by Tseytlin \cite{Ts}.
In summary, there exist  two superinvariants quartic in the
gauge-field strength \cite{Roo},
 which are only distinguished by
the group-index contractions:
$$
I_1 = t_8\  tr F^2 trF^2 - {1\over 4}\epsilon_{10}\  C\  trF^2 trF^2
\ ,
\eqno(14b)
$$
and
$$
I_2 = t_8\  tr F^4 - {1\over 4}\epsilon_{10}\  C trF^4 \ . \eqno(14a)
$$
Here $C$ is the  antisymmetric 2-form,
$\epsilon_{10}$ the Levi-Civita tensor, and
 $t_8$ the
covariant extension of the well-known light-cone-gauge zero-mode
tensor.
The parity-even part of $I_1$ appears however also independently,
in the supersymmetric completion of the  Chern-Simmons-modified
two-derivative action.  
Since all  these superinvariants have anomaly-cancelling pieces
one may  expect that they appear at one given order in the
loop expansion. In heterotic $SO(32)$  theory in ten dimensions
the two-derivative action comes from the sphere,
$I_1$ does not appear at all, while
 $I_2$ appears at one loop only. 
 Duality maps
 the  $\sigma$-model metrics and string couplings as follows
\cite{W}:
$$
\lambda^h = 1/\lambda^I \ ,\ \  G_{\mu\nu}^{h}  =
 G_{\mu\nu}^I/\lambda^I
\eqno(15)
$$
Simple power counting then shows
 that the parity-even parts of $I_2$ and $I_1$
should arise in  type-I theory  from surfaces of Euler number,
 respectively,
minus one  (disk, projective plane)
and one (disk with two holes  etc). This is compatible with
the absence of all  quartic terms at Euler number zero,
as we have concluded.

 Consider now torroidal compactifications.
The gauge-field-dependent
 one-loop free energy in heterotic
$SO(32)$ theory reads \cite{L}
$$
{\cal F}_h^{(1)} =
-{V^{(d)}(\lambda^I)^{4-d/2} \over
2^{10} \pi^6 }\times
  \int_{ Fun}{d^2\tau \over
\tau_2^2}\;{\Gamma^{10-d,10-d}\over (2\pi^2 \tau_2)^{d/2-5}}
A(F,\bar\tau)
\eqno(16)
$$
Here $\Gamma^{10-d,10-d}$ stands for the usual sum over the
Lorentzian
Narain lattice, which factorizes in the integrand because we assumed
 zero Wilson lines,
and
$$\eqalign{
A( F,\bar\tau )=t_8\  tr F^4
+&{1\over 2^9\cdot 3^2} \Bigl[  {E_4^3\over \eta^{24}} +
 {\hat E^2_2 E_4^2\over \eta^{24}}  \cr &
-2 {\hat E_2E_4E_6\over \eta^{24}} -2^7\cdot
3^2\Bigr]\   t_8 ( tr F^2)^2
\cr}
\eqno(17)
$$
with  $E_{2n}$  the $n$th Eisenstein series:
$$
E_{2}=
{12\over i \pi}\partial_{\tau}\log \eta
=1-24\sum_{n=1}^{\infty}{n\, q^n\over 1-q^n}
\eqno(18a)
$$
$$
E_{4}=
{1 \over 2}\left(
{\vartheta}_2^8+
{\vartheta}_3^8+
{\vartheta}_4^8
\right)
=1+240\sum_{n=1}^{\infty}{n^3q^n\over 1-q^n}
\eqno(18b)
$$
$$
E_{6}=
 \frac{1}{2}
({\vartheta}_2^4  + {\vartheta}_3^4 )
({\vartheta}_3^4 + {\vartheta}_4^4 )
({\vartheta}_4^4 - {\vartheta}_2^4 ) 
 =1-504 \sum_{n=1}^{\infty}{n^5q^n\over 1-q^n}
\eqno(18c)
$$
The $E_{2n}$'s
 are modular forms of weight $4n$ except for $E_2$ which
must be modified to
$$
\hat E_2=E_2-{3\over \pi\tau_2}
\eqno(18d)
$$
The powers of $\lambda^I$ in front of expression (16) come  from
the fact that we used  type-I normalizations for the metric:
$d/2$ of these powers are due to  the space-time volume, and the
other
four to the tensor $t_8$.
As for the fact that all holomorphic dependence in the
 integral appears  through the sum over  the Lorentzian lattice, this
is a result of  BPS saturation \cite{HM}.
 It can be derived  by an argument similar
to the one for the open string, except that the background field
now only
couples to the  helicity coming from the left (supersymmetric)
sector.

  The Lorentzian lattice  involves a  sum
 over both momenta and windings on the $(10-d)$-dimensional
 torus.
Setting to zero
 the antisymmetric tensor background, which is
 a Ramond-Ramond field in type-I theory, and using again
type-I normalizations for the compactification torus, we have
$$\Gamma^{10-d,10-d}  =
  \sum_{p\in^*\Gamma\atop  w\in\Gamma}
 e^{-\pi\tau_2 p^2 \lambda^I/2 - \tau_2 w^2/2\pi \lambda^I + i
\tau_1 p\cdot w} \ .
\  \eqno(19)
$$
Now since inside the fundamental domain,
 $\tau_2$ is bounded away from the origin,
all terms with non-zero winding are  non-perturbatively small
at weak $\lambda^I$. This is consistent  with the fact that winding
heterotic strings are solitonic D-branes on the type-I side
\cite{WP}.
The remaining momentum lattice  can   be Poisson-resummed
back and written as follows:
$$
{\Gamma^{10-d,10-d}\over (2\pi^2\tau_2)^{d/2-5}}
 =  {V_{\Gamma}}(\lambda^I)^{d/2-5}\times 
 \sum_{{\tilde w}\in\Gamma}
 e^{- {\tilde w}^2/2\pi  \lambda^I\tau_2}
 + \ o(e^{-1/\lambda^I})
\ .\eqno(20)
$$

We will now plug  the above expression
  into eq. (16), and perform the modular integration.
The ${\tilde w}=0$ term can be
integrated explicitly, using the formulae
$$\eqalign{&\ \ \ \
{I(0,0,0)}=\pi/3\;\;\;,\;\;\;{I(1,1,1)}=-48\pi
\cr &\ \ \ \
{I(0,3,0)}=240\pi\;\;\;,\;\;\;{I(2,2,0)}=48\pi \cr}
\eqno(21a)
$$
where we defined
$$I(m,n,k)=\int_{ Fun} {d^2\tau\over
\tau_2^2}\;{\hat E_2^{m}E_4^nE_6^k\over \eta^{24p}}
\eqno(21b)
$$
subject to the  modular-invariance condition  $6p=m+2n+3k$.
In what concerns the ${\tilde w}\not= 0$ terms, we may extend
their integration regime to the
 entire strip $-{1\over 2} < \tau_1 <{1\over 2}$, modulo
 a non-perturbatively small
error. This makes  the $\tau_1$ integration straightforward, since
only terms without  exponential
$\bar \tau$-dependence in  $A(F,\bar\tau)$ survive:
$$
\int_{-{1\over 2}}^{1\over 2} d\tau_1 A(F,\bar\tau) =
 tr F^4
  + {1\over 8} (tr F^2)^2 \times
 \Bigl[ 1 -{15\over 2\pi\tau_2} + {63\over 8\pi^2\tau_2^2}
\Bigr]
\eqno(22)
$$
Putting all this together,  redefining $\tau_2 \equiv 1/l\lambda^I$,
and doing some tedious algebra leads to our final expression for
the  heterotic one-loop free
energy at weak type-I coupling::
$$\eqalign{
{\cal F}_h^{(1)} =
 -{V^{(10)}\over 2^{10}\pi^6}\ \Biggl\{ &
t_8\; tr F^4\;
 \left( {\pi\over 3\lambda^I} + \int_0^\infty dl\  {\cal K}
\right) +
 \cr
+ {1\over 8}t_8 & \left( tr F^2  \right)^2  \int_0^\infty dl
\  {\cal K}\
\left(1 - {15l \over 2\pi}\lambda^I
 +{63 l^2\over 8\pi^2} (\lambda^I)^2
\right)
+  o(e^{-1/\lambda^I}) \Biggr\}
\cr}
\eqno(23a)
$$
where here
$$
{\cal K} = \sum_{{\tilde w}\in\Gamma-\{0\}} e^{-{\tilde w}^2 l/2\pi}
\ . \eqno(23b)
$$

The leading,  $o({1/\lambda^I})$ term in this expression
corresponds to  the type-I  disk-diagram \cite{Ts}.
The constant  piece should be compared to the sum of
 the  M\"obius-strip and annulus, given by  eq. (13).
These are  indeed identical,  if one notes
 that for a simple  magnetic field
$t_8 F^4 = 24 B^4$. The  remaining terms, as well as the
moduli-independent contribution of the heterotic sphere-diagram
\cite{Ts}
$$ {\cal F}_h^{(0)} =
 { V^{(10)}\over 2^{10}} \ \lambda^I\
t_8\; (tr F^2)^2  \eqno(24)
$$
 correspond to  two- and three-loop diagrams on the
type-I side. If we assume  exact duality and
 no further corrections on the heterotic
side, we conclude  that beyond three loops there are only
instanton corrections on the type-I side.

{\it Afterword}. The effective expansion parameter in eq. (23)
is $ \alpha^\prime_h/ R^2$, where $\alpha^\prime_h  = \lambda^I
\alpha^\prime_I$ is the heterotic Regge slope, and $R$ is a
typical radius of the compactification torus.
Stretched heterotic strings are (non-perturbative)
charged BPS states on the type-I side, so it  is
 not surprising that they should  control at least
part of the   $F^4$ terms in  ${\cal L}_{eff}$.
The role of analogous  degrees of freedom,
as well as of  the two-loop
renormalization, eq. (24),  in the D-brane context must be
elucidated further. The study of fundamental-string
 scattering \cite{pre} may  shed some different
 light on these issues.

\centerline{\bf Acknowledgements}

We  thank the organizers of the Spring
Workshop on String Theory for the invitation.
C.B. aknowledges support from  EEC contract CHRX-CT93-0340,
and thanks  M. Green, S. Shenker, A. Tseytlin
 and P. Vanhove for conversations
on some related issues.

\end{document}